# LearnWeb-OER: Improving Accessibility of Open Educational Resources

Jaspreet Singh
Forschungszentrum L3S
Appelstr. 9a
30167 Hannover, Germany
+49 511 76217713
singh@l3s.de

Zeon Trevor Fernando
Forschungszentrum L3S
Appelstr. 9a
30167 Hannover, Germany
+49 511 76217713
fernando@l3s.de

Saniya Chawla
IIM Ahmedabad
Vastrapur
Ahmedabad, Gujarat, India
+91 79 6632 3456
chawlasaniya21@gmail.com

## ABSTRACT

In addition to user-generated content, Open Educational Resources are increasingly made available on the Web by several institutions and organizations with the aim of being re-used. Nevertheless, it is still difficult for users to find appropriate resources for specific learning scenarios among the vast amount offered on the Web. Our goal is to give users the opportunity to search for authentic resources from the Web and reuse them in a learning context. The LearnWeb-OER platform enhances collaborative searching and sharing of educational resources providing specific means and facilities for education. In the following, we provide a description of the functionalities that support users in collaboratively collecting, selecting, annotating and discussing search results and learning resources.

**Track**: Open Track

## 1. INTRODUCTION

Even though OERs are multiplying on the Web and plenty of data is published online with the aim of being re-used, it is still difficult to find relevant learning resources since they are not uniformly classified and organized, and there are no clear interfaces and tools that facilitate searching. Several projects (e.g. http://linkedup-project.eu/) are committed to improving this situation, providing interoperability and the exploitation of the resources already available on the Web through Linked Data principles and catalogues. Creating a global catalogue is not enough though. What is required is a search interface that helps non-technical users in searching and finding materials useful for learning in different scenarios (at school, at work, or in their private life).

The contribution of the LearnWeb-OER system is to help educators and students in finding relevant educational materials for specific learning and teaching scenarios [2, 5]. Thanks to the feedback we collect from real users, we focus on their needs and requirements to design an interface that facilitates search and collaborative learning. For example, we integrated the TED talks videos (www.ted.com) along with all transcripts and several learning activities to support language learning and teaching. We differentiate ourselves from the official TED website's search feature by allowing users to search for words and phrases within the transcript.

## 2. SYSTEM DESCRIPTION

LearnWeb-OER provides users with a single platform to help them attain their learning goals. It can be seen as part search engine, part social network and part repository. LearnWeb-OER's major features are *Searching* (Web2.0 sources and Linked Open Data datasets are queried through a unified search API), *Sharing* (users can add resources into groups directly from the search page or by uploading it from one's desktop in order to share them with other LearnWeb-OER users), and *Analysis* (users can add comments and additional metadata to enrich the resources).

The web application uses Java and Java Server Faces 2 as web framework. A purely Java based approach was chosen to allow easy extension of the system and good scalability. LearnWeb-OER currently supports the following datasets through its search API:

- Web2.0 user generated contents from YouTube, Vimeo, Slideshare, Flickr
- Web: Bing
- Linked Data: TED Talks Linked Open Data dataset

Web datasets are accessed using the provided API, while the TED dataset is accessed via its SPARQL endpoint. The different datasets that the platform exposes to the user can be accessed through a universal search interface. The interface allows for standard keyword search and uses a rudimentary aggregated search approach to rank results from the various datasets. The TED dataset was created at L3S (www.l3s.de) in 2014 and was recently integrated in LearnWeb-OER. The dataset consists of more than 1800 talks along with a rich collection of 35000+ transcripts present in over 100 different languages at the time of writing. To keep the TED videos up to date, we use a simple crawler script that adds new TED videos to the dataset. The crawler also adds any new transcripts of previously existing videos. The development team controls the crawler manually. The last crawl was in June 2014. Some initial information about the endpoint is described at http://datahub.io/dataset/ted-talks.

The TED Talks dataset includes two main conceptual types: metadata of the actual talks (obtained via the TED API) and additional transcripts for each talk in multiple languages, where the latter is provided as part of our work. We rely on established vocabularies and schemas, most notably the Bibliographic Ontology (BIBO), the Ontology for Media Resources, Dublin Core and schema.org, where talks are represented as bibo:AudioVisualDocument, and transcripts as bibo:Document.

The platform also includes learning activities using the transcripts of the TED dataset. The interface consists of a two columns layout in which users can watch the video as well as follow the related transcript. Users have the option of highlighting a word or part of a sentence in the English transcript and annotating it. When a single word is selected, the platform returns a set of definitions for each relevant Part of Speech (PoS) for that word along with a few related synonyms. The definitions and the synonyms are retrieved from the WordNet lexical database using the RiTa Wordnet Java API (http://www.rednoise.org/rita); this information is then displayed as a tooltip for the highlighted word. Semantic similarity between the word highlighted and the set of synonyms retrieved from the database is calculated by combining the local context and WordNet similarity [4] for adverbs, adjectives and by considering the extent of overlap of the dictionary definition [1] for nouns and verbs. An online demo of LearnWeb_OER is available at: http://learnweb.l3s.uni-hannover.de/linkedup/lw/index.jsf

The LearnWeb-OER platform provides search over the TED talks based on the title, description and from within the transcripts provided for each TED talk. The TED videos are displayed in the search interface along with content from other video sources such as YouTube and Vimeo. The preview of a TED video displays snippets from the transcript where the searched keywords occur. From this preview, users can launch the learning activity designed for TED videos.

## 3. DISTINGUISHING FEATURES

**Innovation in Education:** The goal of LearnWeb-OER is to foster active participation and creativity among students. Searching becomes part of the larger process: the social networking layer helps creating a sense of awareness and users' knowledge. Learning activities based on the integrated datasets (e.g. TED talks) are increasingly provided in close collaboration with our users. All activities are tracked so that teachers can monitor students and convey further instructions through an internal messaging module.

**Audience:** LearnWeb-OER current target users are teachers, trainee teachers and students who want to learn or improve a foreign language using authentic materials in a participatory experience. Several additional learning scenarios can be supported by the diverse features already provided.

**Usability:** The front end is developed with close adherence to user interface design and usability principles. The search interface allows users to access and display various sources, as well as linked data within the same layout. A QUIS survey [3] based evaluations carried out in 2012 and 2013 showed high levels of user satisfaction.

**Performance:** The search interface is comparable to the ones of Google and Bing, it supports infinite scrolling and shadow boxes for full previews of resources. The major difference compared to other search engines is that LearnWeb-OER indexes both the collected resources and the corresponding metadata (tags, comments and ratings) included by users in the various groups. For all resources, we store icons and metadata from the original source, as well as metadata added by LearnWeb-OER users

**Data usage and quality:** We focus on high quality sources covering various types of multimodal resources both from the Web (user-generated-content for education) and from the linked data cloud (e.g. TED dataset). . Users interact with the resources shared in groups by leaving comments or tags. We also keep track of a user's search behavior in high detail.

Some usage statistics are reported in table 1.

**Legal & Privacy**: Users access the platform through login and password. We collect activity logs and statistics only for research purposes. These data are not available to users. To carry out system evaluations we ask users to sign a Research Consent Form in full compliance with privacy and anonymity principles.

| Number of Users | 3358 |
|---|---|
| Number of Groups | 540 |
| Number of Resources Saved | 9679 |
| Number of Youtube Videos | 2117 |
| Number of Vimeo Videos | 417 |
| Number of Flickr photos | 529 |
| Number of Courses | 34 |
| Number of Comments | 1449 |

Table1. Data usage statistics as of 08-09-2014

## 4. CONCLUSION

LearnWeb-OER provides an integrated environment to search over various datasets including Web, Web2.0 and Linked Data sources with a strong focus on ease of use and support of the whole search process including grouping, sharing and annotation of resources. The TED dataset is already a valuable resource for the language learning community, and other suitable datasets will be integrated in the future.

## 5. ACKNOWLEDGMENTS

LearnWeb-OER is founded by internal funds by the L3S Research Center, University of Hannover, Germany. A special thanks to Philipp Kemkes who contributed in the development of the LearnWeb-OER system from the beginning.